\def\spose#1{\hbox to 0pt{#1\hss}}

\def\multleft#1{\hbox to size{\vbox {\halign {\lft{##}\cr #1}}\hfill}\par}
\def\multright#1{\hbox to size{\vbox {\halign {\rt{##}\cr #1}}\hfill}\par}

\def\degmark{^\circ}
 \def\today{\ifcase\month\or January\or February\or March\or April\or May\or
June\or July\or August\or September\or October\or November\or December\fi
\space\number\day, \number\year}

\def\Msun{\hbox{$\rm\thinspace M_{\odot}$}}
\def\Msph{\hbox{$\thinspace M_{{\rm sph}}$}}
\def\Mbh{\hbox{$\thinspace M_{{\rm bh}}$}}

\newcommand{\Msolar}{\mbox{\,$\rm M_{\odot}$}}

\def\H2{\hbox{H$_{2}$}}

\newcommand{\gtsim}{\mbox{{\raisebox{-0.4ex}{$\stackrel{>}{{\scriptstyle\sim}}
$}}}}
\newcommand{\ltsim}{\mbox{{\raisebox{-0.4ex}{$\stackrel{<}{{\scriptstyle\sim}}
$}}}}
\documentclass{mn2e}

\usepackage{times}
\usepackage{amssymb}
\usepackage{epsfig}

\begin{document}
\hsize=6truein
          
\title[On the evolution of the black hole:spheroid  mass ratio]
{On the evolution of the black hole:spheroid mass ratio}

\author[R.J.~McLure et al.]
{R. J. McLure$^{1}$\thanks{Email: rjm@roe.ac.uk}, M.J. Jarvis$^{2}$, 
T.A. Targett$^{1}$, J. S. Dunlop$^{1}$ and P. N. Best$^{1}$\\
\footnotesize\\
$^{1}$Institute for Astronomy, University of Edinburgh, 
Royal Observatory, Edinburgh, EH9 3HJ, UK\\
$^{2}$Astrophysics, Department of Physics, Keble Road, Oxford OX1 3RH}
\maketitle

\begin{abstract}
We present the results of a study which uses the 3CRR sample of 
radio-loud active galactic nuclei to investigate the evolution
of the black-hole:spheroid mass ratio in the most massive early-type 
galaxies from $0<z<2$. Radio-loud unification
is exploited to obtain virial (line-width) black-hole mass estimates
from the 3CRR quasars, and stellar mass estimates from the 3CRR radio
galaxies, thereby providing black-hole and
stellar mass estimates for a single population of early-type galaxies. 
At low redshift ($z\ltsim1$) the 3CRR sample is consistent
with a black-hole:spheroid mass ratio of $M_{bh}/M_{sph}\simeq0.002$, in
good agreement with that observed locally for quiescent galaxies of similar
stellar mass ($\Msph\simeq 5\times10^{11}\Msun$). However, over the
redshift interval $0<z<2$ the 3CRR black-hole:spheroid mass ratio is
found to evolve as $M_{bh}/M_{sph}\propto
(1+z)^{2.07\pm0.76}$, reaching $M_{bh}/M_{sph}\simeq0.008$ by 
redshift $z\simeq2$. 
This evolution is found to be inconsistent with the local 
black hole:spheroid mass ratio remaining constant at a moderately 
significant level ($98\%$). If confirmed, the detection of evolution in the 
3CRR black-hole:spheroid mass ratio further strengthens the evidence 
that, at least for massive early-type galaxies,
the growth of the central supermassive black hole may be completed before
that of the host spheroid.
\end{abstract}


\section{INTRODUCTION}

It is now established at low redshift that black-hole 
mass and host spheroidal mass are tightly correlated in both quiescent 
(Magorrian et al. 1998; Gebhardt et al. 2000; Ferrarese \& Merritt
2000) and active galaxies (McLure \& Dunlop 2001,2002; Nelson et
al. 2004). Consequently, it is widely accepted that the formation and
evolution of supermassive black-holes and their host spheroids must be
intimately related. However, at present we do not have a good
understanding of the origins of the black-hole:spheroid connection, or
its evolution with redshift. Fundamentally, it is not known whether
black-holes formed before, after or co-evally with their host spheroids.
From a theoretical perspective, most models (eg. Granato et
al. 2004; Hopkins et al. 2005; Begelman et al. 2005) predict that 
the $\Mbh-\sigma$ relation should remain close to the locally 
observed relation at all epochs. If true, then the same models predict
that the observed ratio between black hole and spheroid mass should
evolve with redshift as $M_{bh}/M_{sph}\propto(1+z)^{1.5-2.5}$
(e.g. Wyithe \& Loeb 2003; Adelberger \& Steidel 2005). 

Within this
context, several recent studies have attempted to explore the redshift
evolution of the relationship between black-hole mass, spheroid mass
and stellar-velocity dispersion from an observational perspective.
Using the width of the [O{\sc iii}]
emission line as a proxy for stellar-velocity dispersion, Shields et
al. (2003) found no evidence that high-redshift quasars ($z\ltsim3$) 
deviated from the local $\Mbh-\sigma$ relation (Tremaine et al. 2002), 
although the uncertainties were large. Similarly, Adelberger \& Steidel (2005)
examined the quasar-galaxy cross-correlation function of a sample of
79 $z\simeq2.5$ quasars, and found them to be consistent with the
local black-hole:halo mass correlation of Ferrarese (2002),
although again the uncertainties were large. In contrast, based on 
stellar-velocity dispersion measurements for a small sample of Seyfert 
galaxies, True et al. (2004) concluded that at $z\simeq0.4$, 
black-hole masses are $\Delta \Mbh=0.16\pm0.12$ dex larger than 
observed locally, for a given stellar-velocity dispersion. 

In terms of the relationship between black-hole and spheroid mass, 
Peng et al. (2005) recently concluded that the $\Mbh:\Msph$ ratio was a 
factor of $3-6$ times larger at $z\simeq2$ than the present day, based on a 
sample of 15 quasars. Furthermore, at the very 
highest redshifts, there are also indications
of evolution in the $\Mbh-\Msph$ relation. The highest redshift quasar
currently identified ($z=6.41$) by the Sloan Digital Sky Survey 
(SDSSJ1148+5251; Fan et al. 2003) is believed to 
harbour a central black-hole of mass
$\simeq3\times10^{9}\Msun$ (Willott, McLure \& Jarvis 2003; Barth et
al. 2003). However, using CO observations, Walter et al. (2004)
recently estimated a dynamical mass of only $\simeq 5\times
10^{10}\Msun$ for SDSSJ1148+5251, clearly inconsistent with the locally
observed $\Mbh-\Msph$ ratio of $\simeq 0.002$ (Marconi \& Hunt 2003;
H\"aring \& Rix 2004).


The large uncertainties associated with the results discussed above
illustrate that addressing the evolution of the
black-hole:spheroid relation from an observational perspective is
still a difficult proposition. The reasons for this are two-fold.
Firstly, although in principal it is possible to obtain relatively 
accurate spheroid mass estimates for quiescent galaxies out to high 
redshifts, at present no viable technique is available to 
estimate their central black-hole mass. Secondly, although it is now 
possible to obtain black-hole mass estimates for broad-line active
galactic nuclei (AGN) out to the
highest redshifts using the so-called virial technique 
(Wandel, Peterson \& Malkan 1999; Kaspi et al. 2000; McLure
\& Jarvis 2002; Vestergaard 2002), it remains extremely challenging to 
obtain accurate spheroid mass estimates for statistically significant 
samples of luminous high-redshift quasars.

In this paper we explore the possibility of exploiting the unification
of radio-loud AGN to provide spheroidal and black-hole mass 
estimates for a {\it single} population of massive
early-type galaxies in the redshift range $0<z<2$. In the standard radio-loud 
unification scheme (e.g. Barthel 1989; Urry \& Padovani 1995) radio-loud 
quasars and radio galaxies are drawn from the same parent population,
with their classification as either quasars or radio galaxies
dependent solely on whether our line-of-sight to the central engine 
intersects the obscuring torus, or not. Consequently, by using 
a complete sample of low-frequency selected radio-loud AGN we can adopt 
spheroidal mass estimates from the radio-galaxy
component (where the central nucleus is obscured) and black-hole mass
estimates from the quasar component (where we have a direct line of 
sight to the central nucleus) as representative of the whole population. 

The structure of this paper is as follows. In section two we briefly
describe the 3CRR sample. In sections three and four we describe how
the stellar and black-hole masses were estimated. In section five we
investigate the evolution of the $\Mbh-\Msph$ relation within the 3CRR
sample. In section six the implications for the 3CRR $K-z$ relation
are outlined, while in section seven the possible systematic
uncertainties influencing the stellar and black-hole mass 
estimates are reviewed. Our conclusions are discussed in section eight. 
Throughout the paper we adopt the following cosmology: $H_{0}=70$
km\,\,s$^{-1}$Mpc$^{-1}$, $\Omega_{m}=0.3$, $\Omega_{\Lambda}=0.7$.

\section{The 3CRR Sample}

The sample utilised in this study is the 3CRR sample 
(Laing, Riley \& Longair 1983) which contains the most luminous,
low-frequency selected, radio sources in the Northern Hemisphere. The
3CRR sample is complete to a 178MHz flux limit of 10.9Jy over an area
of 4.2 steradians, and contains a total of 173 sources within the
redshift range $0.0<z<2.0$. Following Willott et al. (2003) we have
excluded the nearby starburst M82 (3C231), and two flat-spectrum
quasars which only feature in the 3CRR sample due to Doppler boosting (3C345
and 3C454.3), leaving a total of 170 sources in the full sample.

The 3CRR sample is ideally suited to the purposes of this study 
because it is both complete, and low-frequency selected. The low-frequency 
selection is particularly important because it ensures
that the 3CRR sample does not suffer from any orientation
bias. Indeed, the 3CRR quasar sample is known to be consistent
with being randomly distributed with respect to our line of sight,
with orientation angles restricted to $\leq53\degmark$ (Willott et
al. 2001). In contrast, the quasar components of high-frequency
selected radio samples are known to be dominated by flat-spectrum sources 
which are preferentially aligned close ($\ltsim10\degmark$) to the 
line of sight (Jackson \& Wall 1999). Given the known correlation
between orientation and the width of low-ionization broad emission lines in 
radio-loud AGN (e.g. Wills \& Browne 1986; Brotherton et al. 1996; 
Jarvis \& McLure 2005, in prep), it is therefore likely that virial 
black-hole mass estimates for flat-spectrum quasars are 
systematic underestimates (Jarvis \& McLure 2002). 

Finally, the extensive
optical and near-infrared imaging of the 3CRR sample in the literature has
demonstrated that the radio-galaxy hosts are consistent with being
drawn from the most massive early-types in existence out to $z\simeq
2$ (e.g. McLure et al. 2004; Bettoni et al. 2003; Dunlop et al. 2003;
Best, Longair \& R\"{o}ttgering 1998). Although radio-loud AGN of the
luminosity of the 3CRR
sample are extremely rare, it is worth remembering that the likely radio-source 
lifetime of $\simeq 10^{8}$ years is only a small percentage of
the Hubble time. Furthermore, given that every massive early-type
galaxy must presumably pass through an active phase in order to
build-up its central supermassive black hole, any information
extracted on the evolution of the $\Mbh:\Msph$ ratio from the 3CRR
sample can presumably be applied to the evolutionary history of all
massive early types.

\section{Stellar masses}
It has been known for more than twenty years that powerful radio galaxies 
display a tight correlation between redshift and apparent $K-$band 
magnitude; the so-called $K-z$ relation or near-infrared Hubble 
diagram (Lilly \& Longair 1984). In its simplest interpretation the 
3CRR $K-z$ relation 
is fully consistent with the passive evolution of an $\simeq 3L^{\star}$ 
elliptical galaxy formed in a single burst
of star-formation at $z_{f}\gtsim5$ (e.g. Jarvis et al. 2001; 
Willott et al. 2003). Alternatively, the $K-z$
relation is also consistent with the hierarchical merging paradigm,
within which the $K-z$ relation is populated at each redshift by 
early-type galaxies at a specific stage of their evolutionary history 
(eg. Best, Longair \& R\"{o}ttgering 1998). In either case, the stellar 
populations of the 3CRR radio galaxies ($K-$band luminosities and 
colours) are consistent with a high redshift of formation. 

Consequently, we have based our stellar mass estimates for the 3CRR
radio galaxies largely on the $K-$band photometry compiled by Willott
et al. (2003) for their study of the $K-z$ relation
(corrected to a fixed 64 kpc metric aperture and for emission-line 
contamination). For the purposes of this study we have
restricted ourselves to 3CRR radio galaxies with $z\geq
0.3$. The reasons for this decision are three-fold. Firstly, the lowest
redshift in the quasar component is $z=0.305$, so it is
only at $z>0.3$ that a direct comparison between the radio galaxy and
quasar components of the 3CRR sample can be made. Secondly, by restricting 
our analysis to $z\geq0.3$, the 3CRR radio galaxies all have radio
luminosities higher than the FRI/FRII break (Fanaroff \& Riley 1974),
ensuring that our radio-galaxy sample is comprised entirely of 
powerful FRII sources, the same radio morphology common to the 
entire 3CRR quasar sample. Thirdly, at $z\geq0.3$ the radio
galaxies all have radio luminosities greater than
$L_{151\rm{MHz}}=10^{26}$WHz$^{-1}$sr$^{-1}$, the radio luminosity above
which the 3CRR sample is fully consistent with simple
orientation-based unification (Willott et al. 2001)

The full 3CRR sample contains 57 radio galaxies in the redshift
interval $0.3<z<1.8$. However, in this study we restrict our analysis
to the complete sub-sample of 43 radio galaxies with
$0.3<z<1.8$ and $\delta<55\degmark$, making it possible to compile a
complete set of $K-$band photometry. The photometry for the vast
majority of the sample (39/43 objects) is drawn from the compilation of
Willott et al. (2003). In addition we include $K$-band photometry for 
3C324, 3C356 (Best, Longair \& R\"{o}ttgering 1997), 3C225B (Lilly \&
Longair 1984) and 3C322 (Targett et al., in preparation). 

In the redshift range $0.3<z<1.4$ the stellar masses of the 3CRR
radio-galaxy hosts have been estimated by simply matching the 64 kpc metric
aperture $K-$band magnitudes with a Bruzual \& Charlot (2003) stellar 
population model which undergoes an instantaneous burst of star
formation at a formation redshift of $z_{f}=10$.
At redshifts of $z \ltsim 1$, simply converting the observed $K-$band
magnitudes directly into stellar mass estimates should be relatively robust,
given that the underlying old stellar population should dominate rest
wavelengths long-ward of $1\mu$m. Moreover, there is now a wide 
variety of evidence that at $z\ltsim1$ the properties of powerful 
radio-galaxy hosts (colours,
scalelengths, Kormendy relation) are fully consistent with those of
passively evolving early-type galaxies (McLure \& Dunlop 2000; Best,
Longair \& R\"{o}ttgering 1998; McLure et al. 2004).

However, at $z\gtsim1$ it could be argued that the $K-$band will 
be increasingly contaminated by any on-going star-formation. 
In order to quantify the level of on-going star-formation present in 
the 3CRR radio-galaxy hosts would require a level of multi-wavelength
data which does not presently exist. However, the necessary
multi-wavelength data does exist for the K20 sample
(Cimatti et al. 2004). The K20 sample is comprised of $\simeq500$ 
galaxies in the Chandra Deep-field South region with $K_{vega}<20$, is $97\%$ 
spectroscopically complete, and has {\it ubvrizjhk} photometry available
for each source. Based on this multi-wavelength data-set, Fontana et
al. (2004) conducted a detailed study of the stellar masses of the K20
galaxies using detailed SED fitting. 
Furthermore, using the Fontana et al. results, Daddi et al. (2004)
provide a fitting formula relating observed $K-$band magnitude to 
stellar mass for the K20 galaxies in the redshift range $1.4<z<2.5$, 
the most luminous of which are directly comparable to the 3CRR radio-galaxy 
hosts. Consequently, for the 3CRR
radio-galaxy hosts at $z\geq1.4$ we have adopted the stellar mass
estimates provided by the Daddi et al. (2004) relation derived for the
K20 sample.

The stellar mass estimates for the 3CRR radio-galaxy hosts are plotted
as filled circles in Fig 1. A simple least-squares fit to the $M_{sph}-z$ relation
produces the result:
\begin{equation}
\log(M_{sph}/\Msun)=0.01(\pm 0.06)z +11.73(\pm 0.06)
\end{equation}
\noindent
showing no evidence for evolution out to $z\simeq2$, with the stellar masses 
remaining virtually constant at $\simeq 5\times 10^{11}\Msun$. This places
our stellar mass estimates for the 3CRR sample in good agreement with 
those derived by Rocca-Volmerange et al. (2004) using the {\sc
pegas\'{e}} stellar population models. For completeness we have also
plotted those $z<0.3$ 3CRR radio galaxies with $K-$band photometry
available in Willott et al. (2003). Although they are not included in
our analysis, the $z<0.3$ 3CRR radio galaxies are clearly consistent
with the $M_{sph}-z$ relation fitted to the higher redshift objects.
Finally, it should be noted that we have also chosen to exclude the 20
objects in the 3CRR sample which are classified as weak quasars of
broad-line radio galaxies by Willott et al. (2003). This is
simply because the weak quasars are unsuitable for estimating host
stellar masses (significant point-source contamination) or black-hole
mass (significant host-galaxy contamination). Within the radio-loud 
unification scheme, weak quasars are viewed at
an angle to the line-of-sight intermediate to that of quasars or 
radio galaxies, and their exclusion from our analysis should not 
therefore introduce any significant biases.

\section{Black-hole masses}
\begin{figure}
\centerline{\epsfig{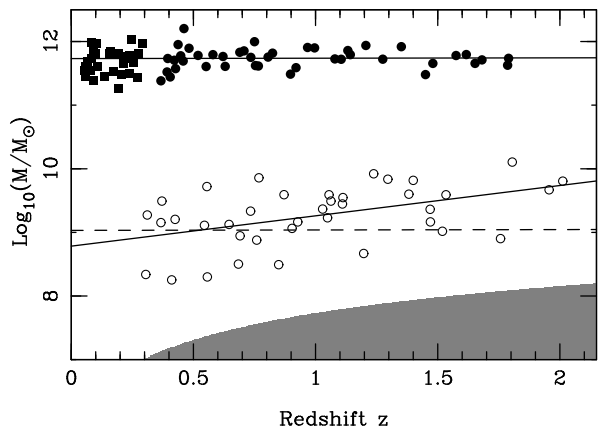}}
\caption{The evolution of the 3CRR radio-galaxy stellar masses (filled
circles) and the 3CRR quasar black-hole masses (open circles) with 
redshift. The solid lines are least-squares fits to the $\Msph-z$ and
$\Mbh-z$ relations (Eqns 1\&\,2). The dashed line shows the predicted 
evolution of the 3CRR quasar black-hole masses under the assumption that the
$M_{bh}:M_{sph}$ ratio remains constant at the local value. The filled
squares show the $z<0.3$ 3CRR radio galaxies which are not included in
the fit to the 3CRR stellar masses (but are consistent with it). The
shaded grey area illustrates the region from which the 3CRR quasar
black-hole masses are excluded by the radio flux limit (see text for a
discussion).} 
\label{fig1}
\end{figure}

The black-hole mass estimates for the 3CRR quasars are based on
the so-called virial black-hole mass estimator for broad-line AGN
(eg. Wandel, Peterson \& Malkan 1998; Kaspi et al. 2000). Under the
assumption that the broad-line emitting gas is in virial motion within
the central black-hole's gravitational potential, the central black-hole mass
can be estimated via: $M_{bh}\simeq G^{-1}R_{blr}V^{2}$, where $R_{blr}$ is
the radius of the broad-line region (BLR) and $V$ is the orbital velocity
of the line-emitting gas. In practise $R_{blr}$ is estimated via
the correlation between optical luminosity and $R_{blr}$ discovered
from reverberation mapping of low-redshift AGN (Kaspi et al. 2000), 
and the gas orbital velocity is taken to be the FWHM of either
the $H\beta$, MgII or CIV emission lines.

Using a combination of literature data and new spectroscopic
observations from the WHT telescope we have compiled line-width 
measurements and black-hole mass estimates for 38/40 of the broad-line
quasars in the 3CRR
sample. Full details of the data compiled on the 3CRR quasars, and the
determination of the black-hole mass estimates are provided in the
Appendix. 

The black-hole mass estimates for the 3CRR quasars are plotted as open
circles in Fig 1. The best fit to the $M_{bh}-z$ relation (solid line)
has the form:
\begin{equation}
\log(M_{bh}/\Msun)=0.48(\pm 0.14)z +8.78(\pm 0.15)
\end{equation}
\noindent
which is inconsistent with no evolution at the $>3\sigma$ level. 
The dashed line in Fig 1 shows the predicted evolution of the quasar 
black-hole masses under the assumption that the black-hole masses
remain a factor of $f=0.002$ lower than the spheroid masses obtained from the
radio-galaxy hosts (Eqn 1). We have adopted $\Mbh/\Msph=0.002$ because
this is the ratio predicted by the recent studies of the local $\Mbh:\Msph$
ratio by Marconi \& Hunt (2003) and H\"{a}ring \& Rix (2004) for a 
spheroidal mass of $5\times10^{11}\Msun$.The dashed
line produces a fit with $\chi^{2}=63.9$ (36 degrees of freedom),
compared to the best-fit with $\chi^{2}=40.6$. For two
free parameters this $\Delta \chi^{2}=23.3$ suggests that the redshift 
dependence of the 3CRR $\Mbh/\Msph$ ratio is different from no
evolution of the redshift zero normalisation at the
$>99.99\%$ \footnote{All significances
based on $\Delta \chi^{2}$ assume Gaussian errors. It should be noted
that significances could be reduced depending on the true form of the
error distributions.} level.

Although the 3CRR sample is purely radio selected, and therefore 
does not have a defined optical flux limit, it is subject to an
{\it effective} optical flux limit through the correlation between radio and
optical luminosity (e.g. Serjeant et al. 1998). Consequently, it is
of obvious concern that the apparent redshift evolution of the 3CRR
quasar black-hole masses may simply be the result of Malmquist bias. 

However, it is relatively straightforward to demonstrate that this is
not the case. As a function of redshift we are interested in
calculating the minimum virial black-hole mass that one of the 3CRR quasars
could have, whilst still producing a radio source more luminous than
the 3CRR flux limit. Therefore, at each redshift we first calculate the
absolute radio luminosity of a source with an apparent flux density
equal to the 10.9Jy 3CRR flux limit (Laing, Riley \& Longair
1983). Secondly, using the optical luminosity - radio luminosity
distribution of the 3CRR quasars (Fig 2) we then determine the minimum
allowable optical luminosity. In Fig 2 the adopted limiting optical
luminosity as a function of radio luminosity is shown as the solid
line. Although this linear relation between optical and radio
luminosity provides a good description of the lower envelope of the
3CRR quasar optical luminosities, it is nevertheless somewhat
arbitrary. However, it can be seen from Fig 2 that moving the adopted limit
to significantly higher optical luminosities is clearly ruled
out. Furthermore, it should be noted that moving the adopted limit to
lower optical luminosities would imply greater sensitivity to lower
black-hole masses, making Malmquist bias less of a concern. Finally,
to calculate the minimum 
virial black-hole mass we assume that the broad emission-line velocity width
must be $>2000$ km s$^{-1}$, if the source is to be regarded as a
broad-line Type 1 quasar. As a function of redshift, the resulting 
minimum allowable black-hole mass, effectively the black-hole mass
completeness limit, is the upper envelope of the shaded grey region
shown in Fig 1.  

The results of this calculation immediately demonstrate that the radio
flux limit of the 3CRR sample does not automatically force the 3CRR 
quasar black-hole masses to increase with redshift. On the contrary,
it remains perfectly possible for the 3CRR quasars to harbour 
black-hole masses as small as $\simeq10^{8}\Msun$ right out to 
redshift $z\simeq2$. Therefore, it seems clear that the redshift
evolution of the 3CRR quasar black-hole masses is genuine, and not the
result of biases inherent to the 3CRR radio selection.

\begin{figure}
\centerline{\epsfig{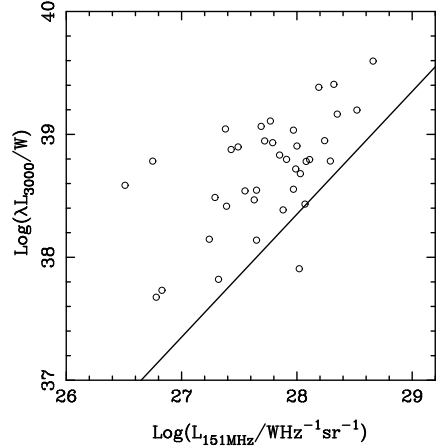}}
\caption{The distribution of the 3CRR quasars on the optical-radio
luminosity plane. The solid line shows the relation for the
lowest likely optical luminosity at a given radio luminosity adopted 
during the calculation of the black-hole mass completeness limit 
(see Section 4).} 
\label{fig2}
\end{figure}

\section{The Evolution of the black-hole:spheroid mass ratio}
\begin{figure}
\centerline{\epsfig{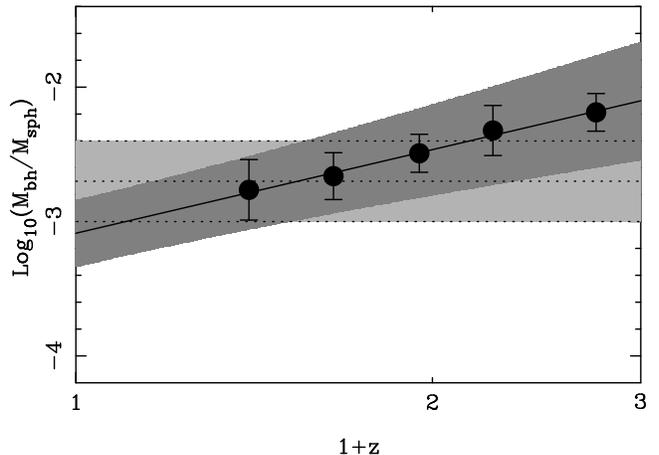}}
\caption{The evolution of the $M_{bh}/M_{sph}$ ratio for
the 3CRR sample (see text for details). The solid line shows the
best-fit to the observed evolution, while the dark grey shaded area shows the
$1\sigma$ uncertainty of this fit. The light grey shaded area
illustrates the $\pm0.3$ dex uncertainty on the local $\Mbh/\Msph$ ratio,
centred on $M_{bh}/M_{sph}=0.002$. The best-fit shown in this figure
has a value of reduced $\chi^{2}\ll1$. However, alternative binning schemes
produce reduced $\chi^{2}$ values consistent with unity, while leaving
the fitted parameters unchanged.}
\label{fig3}
\end{figure}

In Fig 3  we show the evolution of the
$M_{bh}/M_{sph}$ ratio within the 3CRR sample, where each data-point 
is simply $<\log(\Mbh/\Msun)> - <\log(\Msph/\Msun)>$  within each 
redshift bin. The solid line shows the best-fitting relation, which 
has the form:
\begin{equation}
\log(\Mbh/\Msph)=2.07(\pm 0.76)\log(1+z) -3.09(\pm 0.25)
\end{equation}
\noindent
with the dark grey shaded region indicating the $1\sigma$
uncertainty on this fit. It can be seen from Eqn 3 that the slope of 
the best-fitting relation is formally inconsistent with a non-evolving
$\Mbh/\Msph$ ratio at the $2.7\sigma$ level ($99\%$). However, it is
important to note that the expected $\Mbh/\Msph$ ratio for local quiescent
galaxies of the same stellar mass as the 3CRR hosts is still
relatively uncertain. In Fig 3 the light grey shaded area shows
the $\pm0.3$ dex uncertainty on the redshift zero ratio of
$\Mbh/\Msph=0.002$ predicted by the recent $\Mbh-\Msph$
studies of Marconi \& Hunt (2003) and H\"{a}ring \& Rix (2004). It can be
seen from Fig 3 that, although the $\Mbh/\Msph$ ratio of the 3CRR
sample is evolving over the redshift interval $0<z<2$, the
normalisation remains within the current redshift zero uncertainties
until $z\simeq 1.5$. Consequently, in evaluating the significance of
the apparent evolution, it is perhaps better to look for the best
non-evolving fit to the data with a normalisation within the redshift
zero uncertainties (light grey shaded area). Under this constraint,
the best fit to the data has $\log(\Mbh/\Msph)=-2.44\pm0.08$, just
within the redshift zero uncertainties. However, the difference in
chi-square between this non-evolving fit and the best fit is $\Delta
\chi^{2}=7.4$, which for two free parameters is significant at the
$98\%$ level.

Consequently,
we conclude that the data presented in Fig~3 provide strong evidence
that the $\Mbh/\Msph$ ratio within the 3CRR sample evolves with
redshift, changing from $\Mbh/\Msph\simeq0.001$ at redshift zero to
$\Mbh/\Msph\simeq0.008$ by redshift two. However, given the relatively
large uncertainties on the local normalisation of the $\Mbh/\Msph$
ratio, it is also clear that the veracity of the apparent evolution in 
the $\Mbh/\Msph$ ratio of
massive early-type galaxies suggested by the 3CRR sample will only be
thoroughly tested by moving to higher redshifts still.

\section{The near-infrared Hubble diagram}
As discussed previously, the tight $K-z$ relation displayed by the
3CRR sample can either be interpreted in terms of passive evolution or,
alternatively, as fitting naturally within the hierarchical merging
paradigm of galaxy evolution (Best, Longair \& R\"{o}ttgering 1998; McLure \&
Dunlop 2000). In the first interpretation the 3CRR host galaxies
have completed their formation at high redshift ($z\geq2$), and at
all redshifts $z\leq2$ are expected to be indistinguishable from each
other, apart from the passive evolution of their stellar
populations. In the second interpretation the 3CRR host galaxies have
different evolutionary histories, but are seen as powerful radio
galaxies and quasars at a point where their host stellar masses are
a few times $10^{11}\Msun$. In this scenario the 3CRR sources at $z\geq1$
continue to build up stellar mass and are identified with dormant brightest
cluster galaxies at redshift zero.
 
Fundamentally both interpretations rely on the host galaxies of the
3CRR sources having the same mass ($\sim 5\times10^{11}\Msun$), making them
difficult to distinguish via the host-galaxy properties. At $z\leq1$
the results presented in Fig 3 show that the
$\Mbh/\Msph$ ratio of the 3CRR sample is not significantly different 
from the locally observed value, and consequently the passive
evolution interpretation of the $K-z$ relation remains valid for these
objects. Indeed, HST imaging studies have found no evidence that the
luminosities, morphologies or scalelengths of powerful radio galaxies
are evolving out to $z\simeq1$ (e.g. McLure \& Dunlop 2000; McLure et
al. 2004). 

However, although certainly not conclusive, the results presented in
Fig 3 suggest that the 3CRR host galaxies at $z\geq1$ are required to
grow in mass by a factor of $\simeq 2$ to be fully 
consistent with the locally observed $\Mbh/\Msph$ ratio. This level of 
growth would place $z\geq1.5$ 3CRR host-galaxies among the progenitors 
of local 
brightest cluster galaxies ($\Msph\simeq 10^{12}\Msun$), consistent
with the Best, Longair \& R\"{o}ttgering (1998) interpretation of the
$K-z$ relation. Indeed, this scenario is further supported by evidence 
showing that the cluster environments of high-redshift powerful radio 
galaxies are richer than those of their low-redshift counterparts 
(Hill \& Lilly 1991; Best  2000).

\section{Systematic uncertainties}
Before proceeding to discuss the implications of the
suggested evolution in the $\Mbh:\Msph$ ratio for massive early-type
galaxies, it is worthwhile considering the likely sources of
systematic uncertainty.

\subsection{Black-hole masses}
The reliability of the virial black-hole mass estimator for broad-line
AGN has been discussed extensively in the literature (e.g. Krolik
2001; Vestergaard 2002). Perhaps
the most salient point arising from these studies is that comparisons
with the $\Mbh-\sigma$ and $\Mbh-L_{\rm{sph}}$ relations at low-redshift
indicate that the virial mass estimator does not suffer from any
systematic biases (e.g. Onken et al. 2004; Nelson et al. 2004; McLure
\& Dunlop 2002). However, it is worth remembering that while the
virial mass estimator appears to be unbiased, it does carry a
typical $1\sigma$ uncertainty of $\simeq0.4$ dex, and is therefore not
particularly {\it accurate} (Vestergaard 2002; McLure \& Jarvis 2002).

It is therefore of potential concern that more than half of the 3CRR 
black-hole mass estimates shown
in Fig 1 are based on line-width measurements drawn from the
literature (see Appendix). It is possible that this could have an
unforeseen effect on the derived black-hole mass estimates. However,
it appears unlikely that this is the source of the observed evolution
in the 3CRR quasar black-hole masses. Firstly, there is no correlation within
the 3CRR quasar sample between broad-line velocity width and
redshift. Additionally, the scatter apparent in the black-hole mass
estimates plotted in Fig~1 is entirely consistent with that expected,
given the known scatter associated with the virial black-hole mass
estimator (the best-fitting relation shown in Fig 1 has a reduced
chi-square of $\chi^{2}_{red}=1.13$ for 36 degrees of freedom, 
assuming a constant $\sigma=0.4$ dex).

\subsection{Stellar masses}
The stellar mass estimates derived for the 3CRR radio galaxies are in
good agreement with the determinations of Rocca-Volmerange et
al. (2004) and Best, Longair \& R\"{o}ttgering (1998). However, it should
be noted that fundamentally they are based solely on $K-$band
photometry. As previously discussed, at $z\ltsim1$ this should not
significantly bias the stellar mass estimates because the $K-$band
continues to sample rest-frame wavelengths $>1\mu$m, which should be
dominated by the underlying old stellar population. At $z>1.4$ we have
chosen to base our estimates on the correlation between $K-$band
magnitude and stellar mass derived via multi-wavelength SED fitting of
quiescent $1.4<z<2.5$ field galaxies by Daddi et al. (2004). If the
radio-galaxy hosts have significantly less on-going starformation than
typical field galaxies at these redshifts, then these stellar masses
could be underestimates. Indeed, if it is assumed that the radio 
galaxies in the redshift interval $1.4<z<1.8$ are still passively 
evolving, the stellar mass estimates adopted here are 
underestimated by $\simeq40\%$ (e.g. Inskip et al. 2002). It is
important to note that, in this case, the best-fitting evolution of
$\Mbh:\Msph$ ratio becomes $\propto
(1+z)^{1.23\pm0.77}$, and is no longer statistically significant.

In contrast, if the radio-galaxy hosts have
significantly more on-going starformation than typical field galaxies
at these redshifts, then the stellar masses could be
overestimates. Indeed, this is more likely to be the case given the
strong evidence in the literature that AGN activity and starformation
are intimately related at high redshift (e.g. Archibald et al. 2000;
Zirm et al. 2005). However, this possibility is not of great concern to
the present study because, if this is the case, then the apparent
evolution in the $\Mbh:\Msph$ can only become {\it stronger}. Finally,
we note that Spitzer IRAC photometry sampling the rest-frame $K-$band
would be ideal for accurately determining the stellar masses of the
3CRR sample. In this context the results of the current Spitzer 
programme (PI D. Stern) to study the stellar masses of 70 radio galaxies in the
redshift interval $1<z<5$, including several 3CRR objects, will be of
great interest.

\subsection{Unification}
Under a simple orientation-based unification scheme, the ionizing AGN
continuum luminosities of the 3CRR quasars and radio galaxies should
be indistinguishable. Traditionally, the [OIII] emission line
luminosity ($\lambda_{rest}=5007$\AA) has been taken as good 
tracer of the underlying ionizing continuum in radio-loud AGN and, if 
emitted isotropically, should be indistinguishable between quasars and 
radio galaxies. However, studies to directly test the similarity of the [OIII]
emission of radio galaxies and quasars have failed to reach a
consensus, with many finding that the [OIII] emission in quasars is
more luminous than in their radio-galaxy counterparts. The question of
whether this indicates that the [OIII] emission is either non-isotropic 
(or partially absorbed in radio galaxies), or that the opening angle of
the obscuring torus is luminosity dependent, the so-called receding
torus model (eg. Simpson 1998), is still hotly debated and unresolved
(e.g. Grimes et al. 2004; Haas et al. 2005).

Consequently, in this study we have chosen not to make an explicit
receding torus correction to the black-hole mass estimates of
the 3CRR quasars. However, it should be noted
that on average the [OIII] luminosities of the 3CRR radio galaxies are
a factor of $\simeq2$ lower than their quasar counterparts, in good
agreement with the predictions of the receding torus model (e.g. 
Simpson 2003). Correcting the quasar optical luminosities for this
difference would have the effect of lowering their black-hole mass 
estimates by $0.19$ dex (see Appendix). Therefore, although this would
have no effect on the significance of the observed evolution in the
$\Mbh:\Msph$ ratio seen in Fig 3, the five data-points would each be
$0.19$ dex lower.

However, it should also be noted that a systematic increase of only
0.1 dex to the 3CRR radio galaxy [OIII] luminosities is sufficient to
make the [OIII]-$z$ distribution of the quasars and radio galaxies
indistinguishable at the $2\sigma$ level (2DKS test, Peacock
1983). A correction to the 3CRR quasar optical luminosities of this
amount would only change the black-hole mass estimates by $\simeq
0.06$ dex, with a negligible effect on the observed $\Mbh:\Msph$
evolution. Furthermore, recent observations of the 3CRR sample (Haas et
al 2005) suggest that the [OIII] emission in quasars is {\it not}
isotropic, and that the 3CRR sample is fully consistent with
orientation-based unification.
 
\section{Discussion}
\begin{figure}
\centerline{\epsfig{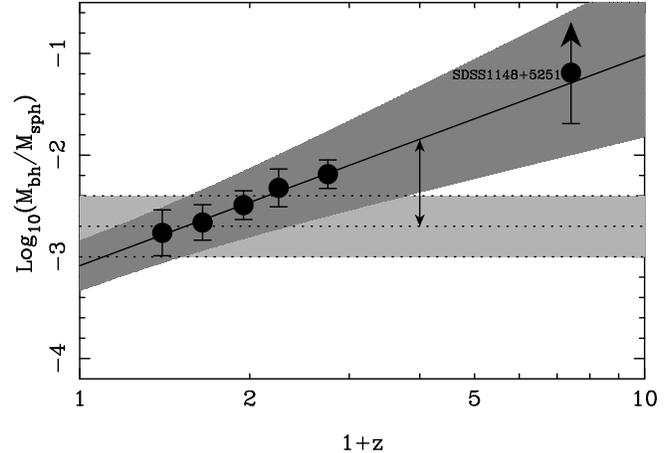}}
\caption{The evolution of the $\Mbh:\Msph$ ratio for the 3CRR sample
as shown in Fig 3, except now extrapolated out to redshift $z=9$. Also
plotted is an estimate of the $\Mbh/\Msph$ ratio for the $z=6.41$ SDSS
quasar SDSSJ1148+5251 (Fan et al. 2003) using the black-hole mass
estimate of Willott et al. (2003) and the dynamical mass estimate of
Walter et al. (2004). The up arrow indicates that the estimated
$\Mbh/\Msph$ ratio is based on the Walter et al. (2004) {\it
dynamical} mass estimate, whereas the enclosed stellar mass could be
considerably lower. The double arrow at $z=3$ illustrates that the $\Mbh:\Msph$
ratio of massive early type galaxies should be a factor of $\simeq 10$
higher by this redshift, if the evolution suggested by the 3CRR sample
is correct.}
\label{fig5}
\end{figure}

The results presented in the last section suggest that the
$\Mbh:\Msph$ ratio within the 3CRR sample is evolving with
redshift, increasing by a factor of $\simeq4$ compared to the local
value by a redshift of $z\simeq2$. However, given that the 3CRR sample is
comprised of the most luminous radio-loud AGN in the Universe, is it 
reasonable to assume that this
result is in anyway representative of the dominant, quiescent
early-type galaxy population? Two lines
of reasoning would argue that it probably is. Firstly, recent studies 
of both the radio-galaxy $K-z$ diagram (e.g. Rocca-Volmerange et
al. 2004) and the host galaxies of powerful radio galaxies (McLure et
al. 2004; Bettoni et al. 2003; Dunlop et al. 2003; McLure \& Dunlop
2000) have demonstrated that 3C-class radio-loud AGN are
representative of the high-mass end of the early-type galaxy
population in terms of  mass, colour, half-light radius and
fundamental plane location. Secondly, as previously mentioned, estimates
of the duty cycle of radio-loud AGN are typically in the range
$10^{7}-10^{8}$ yrs, which accounts for a negligible fraction of the
evolutionary history of the host galaxies ($\simeq 10^{10}$
yrs). Therefore, there seems little evidence to suggest that the
3CRR sample is anything other than a random sample of the very
high-mass end of the early-type mass function over the redshift
interval $0<z<2$. 


Consequently, it is interesting to compare how the results presented
here compare with those in the existing literature. As discussed in
the introduction, from a theoretical perspective the ratio between
black-hole and spheroid mass is expected to evolve as 
$M_{bh}/M_{sph}\propto(1+z)^{\simeq2}$, in excellent agreement with
the results of this study. Furthermore, the evolution of the
black-hole:spheroid relation suggested by the 3CRR sample is also in good
agreement with the observational results of Shields et al. (2003) and
Peng et al. (2005).  However, it is worth noting that there are
further indications from the recent literature which suggest that 
the $\Mbh:\Msph$ ratio may undergo evolution with redshift. Firstly, 
from their study of quasar black-hole masses
using the SDSS first data release, McLure \& Dunlop (2004) concluded
that virtually all black holes in the local Universe with masses
$\Mbh\geq10^{9}\Msun$ were already in place by $z\simeq2$. In
contrast, recent studies of deep infra-red galaxy surveys suggest that
at $z\simeq2$ the mass density comprised of the likely hosts of 
these black holes ($\Msph\geq10^{11}\Msun$) is only $\simeq 30$\% of 
its local value (e.g. Caputi et al. 2005; Fontana et al. 2004; Rudnick et
al. 2003). Qualitatively this would suggest an evolution in the mean
$\Mbh:\Msph$ ratio of a factor of $\simeq 2$.


Furthermore, studies of quasar host galaxies at $z\simeq2$ also
suggest that the $\Mbh:\Msph$ ratio may be
evolving. The majority of recent studies (e.g. Kukula et al. 2001;
Ridgway et al. 2001) suggest that for a given quasar
luminosity, the typical host-galaxy luminosity has dropped by a factor of
$\gtsim 2$ by redshift two. Given that these quasars have similar
nuclear luminosities to those studies at low redshift, and no evidence
has yet been found for a significant correlation between redshift and
quasar broad emission-line widths (e.g. McLure \& Dunlop 2004; Corbett
et al. 2004), it is reasonable to assume that these quasars will have
similar black-hole masses to their low-redshift
counterparts. Consequently, if the host-galaxy luminosities are a
reasonable indicator of mass, it is tempting to conclude that the
detected drop in host-galaxy luminosity implies an increase in the average
$\Mbh:\Msph$ ratio of $\simeq2$. However, it should also be noted that
this conclusion is highly dependent on the unknown star formation
history of the quasar host galaxies, and is as yet based on a very
small number of objects. 

Finally, it is worth returning to the issue of the $\Mbh:\Msph$ ratio
implied for the highest redshift SDSS quasar; SDSS1148+5251
($z=6.41$). Both Willott, McLure \& Jarvis (2003) and Barth et
al. (2003) estimate the central black-hole mass to be $\simeq3\times
10^{9}\Msun$, a value which is consistent with assuming Eddington
limited accretion. In contrast, the dynamical mass estimate of Walter
et al. (2004) is only $\simeq 5\times 10^{10}\Msun$, implying a
$\Mbh:\Msph$ ratio of $\simeq 0.06$, a factor of $\simeq30$ higher 
than observed locally. This is illustrated in Fig 4, which shows the 
implied $\Mbh:\Msph$ ratio of SDSS1148+5251 compared to the 
evolving 3CRR $\Mbh:\Msph$ relation extrapolated to redshift nine. 
Although the good agreement between the two is no doubt fortuitous,
nevertheless, it is clear that the $\Mbh:\Msph$ ratio of
SDSS1148+5251 does not appear to be consistent with the expected local
value. Fig 4 also
demonstrates that the evolution of the $\Mbh:\Msph$ ratio suggested by
the 3CRR sample, if correct, will result in the $z\simeq3$
$\Mbh:\Msph$ ratio being a factor of $\simeq10$ higher than the local
value. Such a significant difference should be detectable directly
from the luminosities of the host galaxies of quasars with virial
black-hole mass estimates. The question of whether the host galaxies
of $z\simeq3$ quasars suggest any evolution in the $\Mbh:\Msph$ ratio
will be addressed in a forthcoming paper (Targett et al., in prep).

\section{acknowledgments}
The authors would like to acknowledge the anonymous referee for
insights which significantly improved the final version of this paper.
The authors would also like to acknowledge Chris Willott for making the
$K-$band photometry for the 3CRR radio galaxies available on-line, and 
Chris Simpson for useful discussions. RJM and PNB would like to 
acknowledge the funding of the Royal Society. MJJ acknowledges the 
funding from a PPARC PDRA. TAT acknowledges the award of a 
PPARC studentship. The WHT telescope is operated on the island of 
La Palma by the Isaac Newton Group in the Spanish Observatorio 
del Roque de los Muchachos of the Instituto de Astrofisica de Canarias

\begin{appendix}
\section{Emission line data}
The adopted broad emission line-widths for 38/40\footnote{No suitable
emission-line data could be found for 3C318. We have chosen to exclude
3C216, often classified as a quasar, because it is known to display
blazar-like properties (Barthel, Pearson \& Readhead 1988), and shows
no indication of 
broad MgII or H$\beta$ in its SDSS spectrum.} of the 3CRR quasars
are listed in Table A1. The emission-line widths for fourteen of the
quasars are derived from new spectroscopic data obtained at the 
William Herschel Telescope, using the ISIS spectrograph, during two 
observing runs in December 2002 and December 2004. Optical
spectroscopy for two further quasars
(3C9 and 3C186) were obtained from the Sloan Digital Sky Survey
(SDSS) archive. The line-widths for these sixteen quasars were
extracted in an automated fashion using the line-fitting software 
originally developed for the analysis of SDSS quasar spectra 
(see McLure \& Dunlop 2004 for a full discussion). The broad 
emission-line widths for the remaining 22 quasars were obtained 
from various literature sources, with the appropriate references 
listed in Table A1. In several cases
where it proved possible to extract the original spectra from the
literature, the emission-line widths listed in Table A1 were
derived using the same line-fitting software employed on the 
new ISIS data.

\subsection{Black-hole mass estimates}
The black-hole mass estimates for the 3CRR quasars were calculated using
the virial mass estimator derived for the MgII emission line by McLure
\& Dunlop (2004):
\begin{equation}
\frac{ M_{bh}} {\Msolar}  =3.2\left(\frac{\lambda
L_{3000}}{10^{37}{\rm W}}\right)^{0.62}\left(\frac{FWHM(MgII)}
{{\rm kms}^{-1}}\right)^{2}
\label{final}
\end{equation}
\noindent
where $\lambda L_{3000}$ is the monochromatic continuum luminosity at
3000\AA. For the 23 quasars with available MgII line-widths
Eqn A1 was applied directly. For the 13 quasars with available 
H$\beta$ line-widths, Eqn A1 was also applied
directly, under the explicit assumption that the line-widths of MgII
and H$\beta$ are, on average, equal. The
reliability of this assumption was the original motivation for
re-calibrating the virial black-hole mass estimator for the MgII
emission line (McLure \& Jarvis 2002), and was demonstrated using
$>1000$ SDSS quasars with optical spectra covering both MgII and
H$\beta$ (McLure \& Dunlop 2004). For the two objects with only CIV
line-widths Eqn A1 was again applied, although the CIV
line-widths were reduced by a factor of $\sqrt{2}$ to account for the
factor of two difference in the emission radius of CIV and H$\beta$
(e.g. Korista et al. 1995); i.e. $V\propto R^{-0.5}$. It should be noted
that this procedure produces virtually identical results to the virial
mass estimator calibration based on CIV derived by Vestergaard (2002).

In addition to an emission line-width, the virial mass estimator relies
on a continuum luminosity measurement. In the majority of cases
(30/38) one or more flux-calibrated spectra were available, from which
the necessary continuum luminosity could be directly measured. For the
remaining quasars it was necessary to calculate the continuum
luminosity from the available $V-$band photometry. This was done by
integrating the composite radio-loud quasar spectrum derived from the
First Bright Quasar Survey (Brotherton et al. 2001) over the $V-$band 
filter curve, normalising the composite spectrum to match the 
available photometry. This normalised composite was then be used to 
predict the continuum luminosity at 3000\AA.

The fact that the data for the 3CRR quasar sample, both line-widths
and continuum luminosities, originates from a combination of new and
literature observations could potentially be a source of increased
scatter in the resulting black-hole mass estimates. Ideally, it
would be preferable to have a complete set of well flux-calibrated
spectra of the 3CRR quasar sample taken with the same instrumentation.
However, despite this, there is no evidence that the black-hole mass
estimates for the 3CRR quasars derived here are any less accurate than
is expected, given the known scatter associated with the virial mass
estimator. Even with superior data quality, previous studies have
determined that the accuracy of the virial mass estimator is limited
to $\simeq0.4$ dex (Vestergaard 2002; McLure \& Jarvis
2002). As discussed in Section 7, the scatter associated with the
$M_{bh}-z$ relation displayed by the 3CRR quasars (Fig 1) is entirely
consistent with this level of scatter.

\begin{table}
\begin{center}
\caption{Details of the 3CRR quasars. Column one lists the source name,
column two lists the source redshifts and column three lists the
$V-$band magnitudes. Column four lists the adopted broad-line FWHM
measurements in units of 1000 km s$^{-1}$ ($\star$ indicates a new line-width measurement based on a published spectrum), while column five lists which
emission line the FWHM measurements were taken from. Column six lists the
logarithm of the derived black-hole mass estimates (units of
$\Msolar$), and column seven provides the
literature references from which the line-width measurements were
taken. Objects listed as ``ISIS'' in column seven are new line-width
measurements based on spectra obtained for this study on the WHT
telescope. Objects listed 
as ``SDSS'' in column seven are new line-width measurements derived 
from publically available SDSS spectra. The numbered references listed
in column seven are as follows: 1. Aars et al. (2005), 
2. Aldcroft, Bechtold \& Elvis (1994), 3. Barthel, Tytler \& Thomson
(1990), 4. Brotherton (1996), 5. Jackson \& Browne (1991), 
6. Lawrence et al. (1996), 7. Marziani et al. (2001)}
\begin{tabular}{lccrccc}
\hline
Source & z&V&FWHM&Line&$M_{\rm{bh}}$&Reference\\
\hline
3C9    & 2.012&  18.2&7.04&MgII&\phantom{0}9.8& SDSS\\

3C14   & 1.469&  20.0&9.68&MgII&\phantom{0}9.4& ISIS\\

3C43   & 1.470&  20.0&6.01&MgII&\phantom{0}9.2& ISIS\\

3C47   & 0.425&  18.1&9.88&MgII&\phantom{0}9.2& ISIS\\

3C48   & 0.367&  16.2&7.32&H$\beta$&\phantom{0}9.2& ISIS\\

3C68.1 & 1.238&  19.5&16.88 &MgII&\phantom{0}9.9&ISIS\\

3C109  & 0.305&  17.9&5.10 &H$\beta$&\phantom{0}8.3&ISIS        \\

3C138  & 0.759&  17.9&5.14&H$\beta$&\phantom{0}8.9&ISIS\\

3C147  & 0.545&  16.9&8.93&H$\beta$&\phantom{0}9.1&6      \\

3C175  & 0.768&  16.6&10.90&MgII&\phantom{0}9.9&ISIS\\

3C181  & 1.382&  18.9&9.82&MgII&\phantom{0}9.6&ISIS\\

3C186  & 1.063&  17.6&8.45&MgII&\phantom{0}9.5&SDSS\\

3C190  & 1.197&  20.0&6.33&MgII&\phantom{0}8.7&1\\

3C191  & 1.956&  18.7&11.30&CIV&\phantom{0}9.7&ISIS\\

3C196  & 0.871&  17.6&9.81&MgII&\phantom{0}9.6&6\\

3C204  & 1.112&  18.2&9.23&MgII&\phantom{0}9.5&ISIS\\

3C205  & 1.534&  17.6&6.40&MgII&\phantom{0}9.6&ISIS\\

3C207  & 0.684&  18.2&3.63&MgII&\phantom{0}8.5&ISIS\\

3C208  & 1.110&  17.4&8.70&MgII&\phantom{0}9.4&ISIS\\

3C212  & 1.049&  19.1&8.58&MgII&\phantom{0}9.2&1\\

3C215  & 0.411&  18.3&4.44&H$\beta$&\phantom{0}8.3&4\\

3C245  & 1.029&  17.3&$\star6.82$&MgII&\phantom{0}9.4&1\\

3C249.1& 0.311&  15.7&7.83&H$\beta$&\phantom{0}9.3&7     \\

3C254  & 0.734&  18.0&9.13&H$\beta$&\phantom{0}9.3&5\\

3C263  & 0.646&  16.3&4.76&H$\beta$&\phantom{0}9.1&7     \\

3C268.4& 1.400&  18.4&$\star$10.66&MgII&\phantom{0}9.8&1\\

3C270.1& 1.519&  18.6&4.50&MgII&\phantom{0}9.0&1\\

3C275.1& 0.557&  19.0&4.40&MgII&\phantom{0}8.3&1\\

3C286  & 0.849&  18.0&$\star$2.25&MgII&\phantom{0}8.5&2\\

3C287  & 1.055&  17.7&$\star$8.73&MgII&\phantom{0}9.6&2\\

3C309.1& 0.904&  16.8&4.23&H$\beta$&\phantom{0}9.1&6\\

3C334  & 0.555&  16.4&10.65&H$\beta$&\phantom{0}9.7&7     \\

3C336  & 0.927&  17.5&7.10&H$\beta$&\phantom{0}9.2&4\\

3C351  & 0.371&  15.3&8.76&H$\beta$&\phantom{0}9.5&7    \\

4C16.49& 1.296&  18.5&$\star$13.98&MgII&\phantom{0}9.8&1\\

3C380  & 0.691&  16.8&4.28&H$\beta$&\phantom{0}8.9&7     \\

3C432  & 1.805&  18.0&11.37&MgII&10.1&1\\

3C454  & 1.757&  18.5&4.78&CIV&\phantom{0}8.9&3   \\
\hline
\end{tabular}
\label{tab1}
\end{center}
\end{table}
\end{appendix}

\end{document}